\documentclass[11pt]{article}

\usepackage{amsthm}
\usepackage{amssymb}
\usepackage{amsmath}

\date {August 16, 2006}
\begin{document}
\author{Sergio Doplicher 
                         \\Dipartimento di Matematica
                         \\University of Rome "La Sapienza"
                         \\00185 Roma, Italy  }

\title{Quantum Field Theory on Quantum Spacetime}
\maketitle

\begin{abstract}
Condensed account of the Lectures delivered at
the Meeting on {\it Noncommutative Geometry in Field and String
Theory}, Corfu, September 18 - 20, 2005.

\end{abstract}

\noindent{\bf Introduction.}
Can Quantum Mechanics and General Relativity merge in a unified consistent
theory? Opinions range from the positive assertion of String Theory to the
opposite extreme; yet a similar question is in a sense still open also for 
Quantum Mechanics and $Special$ Relativity, since no exactly solved
nontrivial model of interacting 
Quantum Field Theory on Minkowski space is yet known. 
However  Quantum Mechanics and Special Relativity do meet in one basic
principle, the $principle$ $of$ $locality$, which by itself (rather
surprisingly, with $unreasonable$ effectiveness) implies most of the
conceptual and formal structure of Quantum Field Theory (e.g., the
existence of a gauge group and the global gauge principle based only on
local $observables$, cf
\cite{h, a}). Do Quantum Mechanics and General Relativity meet in a basic
principle, valid at all scales, which at scales which are large compared
to the
Planck scale reduces to locality, and is equally rigid and fruitful of
structural consequences? Following the lessons of Einstein and Heisenberg,
which marked the birth of Modern Physics, can it be operationally
grounded?

Unfortunately we know too little of Physics at  Planck scale, where
"operational" might loose meaning; yet in this spirit it is possible at
least to draw $limitations$ on the very meaning of Spacetime,
which lead to Quantum Spacetime.
The theory of interacting Quantum Fields on Quantum Spacetime can
then be tackled; the various formulations which are equivalent to
one another on the classical Minkowski space are now distinct; there is no
Osterwalder - Schrader connection between Minkowkian and Euclidian 
Theories, which might well be completey unrelated, in particular as far
as divergences are concerned; causality is lost, as expected, but it is
still unclear whether interaction of fields on Quantum Spacetime is
reconciliable with Lorentz invariance. We summarise merits and
defects of some of the approaches.\medskip

\noindent{\bf The Basic model.}
At large scales spacetime is a pseudo Riemaniann manifold locally modelled
on Minkowski space. But the concurrence of the principles of Quantum
Mechanics and of Classical General Relativity renders this picture
untenable in the small.

Those theories are often reported as hardly reconcilable, but they do meet
at least in a single principle, which concerns just the background
geometry, the $Principle$ $of$
$Gravitational$
$Stability$ $against$ $localization$ $of$ $events$ formulated in
\cite{dfr1, dfr2, dopl1}:
{\bigskip}

{\it The gravitational field generated by the concentration of energy
required by the Heisenberg Un\-cer\-tainty Principle to localise an event
in
spacetime should not be so strong to hide the event itself to any distant
observer - distant compared to the Planck scale.}
{\bigskip}

Already at a semiclassical level, this Principle leads to $Spacetime$
$Un\-cer\-tainty$ $Relations$, that were proposed and shown to be implemented
by
Commutation Relations between coordinates, thus turning Spacetime into
$Quantum$ $Spacetime$ \cite{dfr1, dfr2}. The word "Quantum" is very
appropriate here, to stress that noncommutativity does not enter just as a
formal generalization, but is strongly suggested by a compelling
$physical$ reason (unlike the very first discussions of possible
noncommutativity of coordinates in the pre-renormalisation era, by
Heisenberg, Snyder and Yang, where noncommutativity was regarded as a
curious, in itself physically doubtful, $regularization$ $device$). 

Such an analysis leads to the following conclusions:\medskip

\noindent(i) There is no a priori lower limit on the precision in the
measurement of any $single$ coordinate (it is worthwhile to stress once
more that the apparently opposite
conclusions, still often reported in the literature, are drawn under the
$implicit$ assumption that $all$ the space coordinates of the event are
simultaneously sharply measured); \medskip

\noindent(ii) The Space Time Uncertainty Relations emerging from the
$Principle$
$of$
$Gravitational$ $Stability$ $against$ $localization$ $of$ $events$, in
their weak
form and disregarding the contributions to the source in Einstein
Equations due to the average energy momentum density in generic quantum
states, can be implemented by $covariant$ commutation relations between
the
coordinates, which define a fully Poincare' covariant $Basic$ $Model$ of
Quantum Spacetime. This situation parallels that of Non Relativistic
Quantum
Mechanics, where position and momentum become operators, but they are
acted
upon by the $classical$ Galilei Group.
Indeed the quantum features of spacetime should show up only $in$ $the$
$small$,
at Planck scale, while the Poincare group describes the $global$ $motions$
of spacetime, and should act accordingly the same way both in the small
and in the large, where of course the classical Poincare' group governs
the scene. In particular we can say that spacetime coordinates are
$operators$, but their translation parameters are $numbers$. \medskip   

It is also appropriate to emphasise that we are discussing $Quantum$
$Minkowski$ $Space$, that is we look for quantum aspects in the small
which do not change the large scale flat geometry, in view of a
discussion of interactions between elementary particles. Covariance under
general coordinate transformations is $not$ required. A further step would
be to discuss "operationally" how the general covariance principle ought
to
modify in the Quantum Gravity domain, where, at Planck scale, Einstein
$gedanken$ $experiment$ of the freely falling lift would loose meaning.
\medskip

\noindent(iii) In the Basic Model of Quantum Spacetime, the Euclidean  
distance
between two events and the elementary area have both a lower bound of the
unit order in Planck units; this is quite compatible, as shown by the
model, with Poincare' covariance, and not to be confused with the
unlimited accuracy which is in principle allowed in the measurement of a  
$single$ coordinate.

Similarly, the spectrum of the space 3 - volume operator 
extends down to zero; while the  spacetime 4 - volume operator
is Poincare' invariant,  with pure point spectrum with a
lower bound of the unit order in Planck units \cite{df1}.

These results, in full agreement with the Spacetime Uncertainty Relations,
can be derived from a new interesting interplay of different algebraic,
and
C* - algebraic structures underlying the universal differential calculus.
\medskip

\noindent (iv)  The  Basic Model replaces the algebra of continuous   
functions
vanishing at infinity on Minkowsky Space by a noncommutative C* - Algebra
$ \cal E$,
the enveloping C* - Algebra of the Weyl form of the commutation relations
between the coordinates, which turns out to be the C* - Algebra of
continuous functions vanishing at infinity from  $\Sigma$ to the C* -
Algebra of compact operators. Here $\Sigma$ is the $joint$ $spectrum$
$of$ $the$ $commutators$, which, due to the uncertainty relations, turns
out to be the full Lorentz orbit of the standard symplectic form in 4     
dimensions, that is the union of two connected
components, each omeomorphic to $SL(2,\mathbb C)/ {\mathbb C}_*\simeq
TS^2$.

This manifold does survive the large scale limit; thus, $QST$ $predicts$
$ex\-tra\-di\-men\-sions$, which indeed manifest themselves in a $compact$
manifold
$S^2   \times  \left\{ \pm 1 \right\}$ if QST is probed with $optimally$
$localised$ $states$.
The discrete two-point space which thus appears here as a factor reminds
of the one postulated in the Connes-Lott theory of the Standard Model
\cite{c1, c2, cm}.

In this light QST looks similar to the phase space of a $2-dimensional$
Schroedinger particle; and thus naturally divides into cells (of volume
governed by the 4-th power of the Planck length); so that, though being
continuous and covariant, QST is effectively discretised by its Quantum
nature. (Compare the earlier discussion of the "fuzzy sphere" by John   
Madore). \medskip

\noindent\noindent{\bf Quantum Field Theory.} Quantum Field Theory on the
Basic Model of Quantum Spacetime was
first developed in  \cite{dfr1}; while fully Poincare' Covariant $Free$
Field Theory (as Wightman Fields on QST, or as Poincare' Covariant nets of
von Neumann
Algebras labelled by projections in the Borel completion of $ \cal E $, 
which specify "noncommutative regions" in QST) can be explicitly
costructed, and its violation of causality computed, all attempts to
construct $interacting$ QFT on QST seem to lead sooner or later to   
violations of Lorentz invariance, besides the inevitable violations of
causality.

In the first approach to QFT on QST in  \cite{dfr1}, a natural
prescription was given,
not leading to interacting fields, but to a perturbative expansion of the
S - Matrix, $without$ $manifest$ $violations$ $of$ $unitarity$; but
interaction
required an integration over $ \Sigma$, thus breaking Lorentz invariance
(there is no finite invariant measure or mean on $ \Sigma$), yet
preserving
spacetime translation and space rotation covariance.

The approach based on Yang - Feldman Equation defines perturbatively  
covariant interacting fields, but Lorentz invariance will break a) at the
level of renormalization b) at the level of asymptotic states
\cite{bdfp1, bdfp3}.

Al level a) a fully covariant procedure replaces Wick Products by
$Qua\-si\-pla\-nar$ $Wick$ $Products$ where one subtracts only terms which
are  
$local$ and divergent on QST \cite{bdfp3}; the above problems would
riemerge with a further finite renormalization which ensures to recover 
the
usual renormalised perturbative expansion in the large scale limit. 

A more radical modification of the Wick product is suggested by the very
quantum nature of Spacetime, the $Quantum$ $Wick$ $Product$ \cite{bdfp2}.
It is based on the remark that the usual Wick Product is defined taking
the product
of field
operators at independent points, making the necessary subtractions, and
then taking the limit where all differences of independent coordinates   
tend to zero. But differences of independent coordinates in Quantum
Spacetime are quantum variables themselves, obeying, up to a factor of
order 1, the $same$ commutation relations as the quantum coordinates; and
thus they $cannot$ be set equal to zero. The best one can do is to
evaluate, on functions of several independent quantum coordinates, a
$conditional$ $expectation$ which essentially evaluates on the functions
of each difference
variable an optimally localised state, and leaves the barycenter
coordinates unchanged. Combined with the usual subtractions this procedure
leads to the Quantum Wick Product;
its use to define interactions $regularises$ $completely$ $QFT$ $in$ $the$
$Ultraviolet$;
but Lorentz covariance is broken here by the Quantum Wick Product itself
(while
no integration on $ \Sigma$ is needed here  - the resulting ineraction  
Hamiltonian turns out to be constant on  $ \Sigma$ - the very notion of
optimally
localised state refers to a specific Lorentz frame: it requires that the
sum of the squares of the uncertainties of the four quantum coordinates
is minimal), however, again, spacetime translation and space rotation
covariance are preserved.
Moreover, the Adiabatic
Limit poses serious problems \cite{bdfp2}; see however \cite{dz1, dz2}.  

Thus Lorentz breaking appears in all the above attempts through the
presence of a non trivial centre of the (multiplier algebra of the -) 
Algebra of QST,
whose spectrum is $ \Sigma$: no finite invariant integration is possible
and renormalization introduces a bad dependence on the points of $
\Sigma$.

The problem of Lorenz breaking has received a lot of attention from many  
viewpoints, quite different from the one advocated here (from Quantum
deformation of Poincare' group, notably studied by J.Lukierski,
M.Chaichian, J.Wess and their groups of collaborators, to the recent
discussion of modifications of the coproduct in the Field Algebra
itself, by the same Authors and
T.R.Govindarajan, A. P. Balachandran and others; see other
contributions to this Volume, and compare \cite{z2}. Related is the
discussions of
deformed General Covariance, notably by J.Wess and collaborators, see
the contributions to this Volume by Julius Wess and Paolo Aschieri).
\medskip

But most of these
approaches assume that the commutators of the coordinates are numbers; the
prize to pay is then deformation of the Poincare' group, if not to give up
altogether the principle that laws of nature are the same in all Lorentz
frames - as required in String Theory by the presence of external $ B -
Fields$. \medskip

One might evaluate on QST relativistic interacting quantum fields
(supposedly) already given on Minkowski space; in the perturbative
picture this would amount to evaluate the ordinary, Minkowskian
renormalised interaction density as a function on QST (rather than doing 
this on each field operator). In this case,
the interaction would not be affected by the Quantum nature of Spacetime, 
which would leave no trace at all in the S-Matrix. Some formal
manipulations might just lead to this option.\medskip

A common feature of the S - matrix approaches described above,
based both
on ordinary and Quantum Wick product, is that, while not leading to proper
interacting fields on Quantum Spacetime, they lead, through a natural
ansatz, to a Gell Mann - Low formula for the S - matrix which agrees with
the one which would be derived on $Classical$ Minkowski Space from a $non$
$local$ interaction, obtained from the one given at the start by smearing
with a non local kernel, whose shape reflects the quantum nature of
Spacetime and depends upon the chosen procedure (in the case of the
Quantum Wick product it would be a Dirac delta function of the sum of all
variables, reflecting translation covariance, times a Gaussian in all the
difference variables, which regularises all the Ultraviolet divergences);
such an effective interaction is formally self adjoint and does not lead
to any unitarity violation. \medskip

The crucial point is that Feynmann rules
cannot be freely imposed as a further ansatz, but must be deduced from the
Dyson expansion: the time ordering refers to the times of the
effective interaction Hamiltonians in the ineraction representation, and
$not$ to the time variables in the field operators \cite{dfr1}; thus
Feynmann propagators have to be replaced by the Denk - Schweda
propagators; but full Feynmann rules can be reformulated in a complete way
\cite{p1}. \medskip

\noindent{\bf Other models of Quantum Spacetime. }
Does the commutator of
Quantum coordinates $have$ to be central? This hypothesis, introduced in
\cite{dfr1} on mere simplicity grounds, can be removed in a class of more
elaborate models, where the commutators commute with one another but
not with the coordinates themselves. Yet the centre is again large (and   
not even translation invariant), actually acted upon $freely$ by the  
Lorentz group.

While the model is Poincare' covariant, the irreducible representations  
break (not only Lorentz invariance, as in the basic model, but also)
translation covariance in some direction. The model has however the virtue
of implementing the Spacetime Uncertainty Relations not only in the weak  
form, as reproduced by the Basic Model, but in a somewhat stronger form, 
closer to that suggested by the original semiclassical
argument \cite{bdfp4}.

This feature might imply a better reguralization of interactions: indeed  
already the one implied by the Basic Model for interactions in the
form originally introduced in \cite{dfr1}, has been shown to be sufficient
to reguralise the $ \phi ^{3}$  interaction \cite{b1}; the present  
model might well go beyond.\medskip

\noindent{\bf A New Scenario. } The Principle of
Gravitational Stability ought to be
fully used in the very derivation of Space Time Uncertainty Relations,
which would
then depend also on the energy-momentum density of generic background   
quantum states; this leads to commutation relations between Spacetime
coordinates depending in principle on the metric tensor, and hence, 
through
the gravitational coupling, on the
interacting fields themselves. Thus the commutation relations between
Spacetime coordinates would appear as part of the equations of
motions along with Einstein and matter field Equations.

In other words we may expect that, while Classical General Relativity
tought
us that Geometry $is$ dynamics, Quantum Gravity might show that also
Algebra $is$
dynamics.

This new scenario \cite{dopl2} appears extremely difficult to formalise
and implement, but promises
most interesting developments. Notably it would be related to the   
nonvanishing of the $Cosmological$ $Constant$  \cite{dopl2}  and might
explain
Thermodynamical Equilibrium of the early Universe without Inflation;  it
might
relate to the distribution of correlations in the early Universe shown by
WMAP3.
The key feature of this scenario is that it seems to indicate that      
commutators of
spacetime coordinates, hence the range of acausal effects, $should$ $be$
$larger$
$where$ $gravitational$ $forces$ $are$ $stronger$ and diverge near
singularities. Thus at the very beginning, near the Big Bang, the range of
acausal effects would have been infinite, establishing
thermodynamical equilibrium among well distant regions. Also, the Universe
could have started as an effectively zero dimensional system, and if so
this fact might have left traces in the spectrum of the Cosmic Microwave
Radiation.

Spacetime coordinates and fields would appear as intimately connected
features of the quantum texture of spacetime, which cannot be separated
from one another, exept in some drastic approximation, whose price might
well be the breakdown of Lorentz  invariance of  interactions between
Quantum Fields.

Moreover, QST might teach us something about dark matter if, as expected,
it implies
a $minimal$ $size$ for black holes, where Hawking evaporation would stop.
The minimal black holes would be stable, and fill the Universe with a gas
which would contribute to the dark matter. 

The problem, however, would be displaced to that of their possible
formation \cite{m}\medskip .

\noindent{\bf Phenomenological consequences? }
The $square$ of the Planck length
appears in the commutator of coordinates,  hence Quantum Spacetime
corrections to ordinary interactions are to be expected to be, at the
lowest order, quadratic in the Planck length; if not higher. The effects
are bound to be very small: the difficulties to detect
$directly$ gravitational waves  
(an effect in Classical General Relativity,  for which the Universe
offers sources of tremendously high energy) do not induce to optimism to
see $direct$ effects.

Moreover, as briefly described here above, field interactions on Quantum
Spacetime can well be formulated in a spacetime translation and space   
rotation invariant way, and the lack of a satisfactory formulation which
is also Lorentz invariant may well be just a weak point of the routes
followed so far, rather than an intrinsic impossibility. Therefore it
might well be necessary to look for more subtle tests, rather than
for effects
due to possible spacetime symmetry breakdown.

But the scenario alluded to in the earlier paragraph might well lead to
detect shadows of the Quantum structure of Spacetime in the early
Universe. 

One difficulty is that, as implicit in the discussion of \cite{dfr1}, the
$U(1)$  Gauge Theory (QED) on QST becomes effectively a $U( \infty )$
Gauge Theory and gauge invariant quantities essentially disappear; an
interesting solution  to this difficulty has been proposed by
J.Madore, S.Schraml, P.Schupp and J.Wess, introducing  the concept
of covariant coordinates \cite{mssw}; see also \cite{z1}.

\end{document}